\documentclass[iop]{emulateapj}
\usepackage{amsmath,amssymb,graphicx,natbib}

\newcommand{\mh}{M_\bullet}

\newcommand{\sbh}{SBH}
\newcommand{\sbhs}{SBHs}
\newcommand{\bh}{BH}
\usepackage{url,amsfonts,epsfig}
\usepackage{graphicx}
\usepackage{amsmath}
\usepackage{amssymb}
\usepackage[section]{placeins}
\usepackage{chngpage}
\usepackage{calc}
\usepackage{subfigure}
\usepackage{mathtools}

\usepackage{natbib}
\makeatletter 
\renewcommand\@biblabel[1]{}

\shorttitle{Relativity and the S-star orbits}
\shortauthors{Antonini and Merritt}

\def\gap{\;\rlap{\lower 2.5pt
\hbox{$\sim$}}\raise 1.5pt\hbox{$>$}\;}
\def\lap{\;\rlap{\lower 2.5pt
 \hbox{$\sim$}}\raise 1.5pt\hbox{$<$}\;}

\begin{document}
\title{Relativity and the  evolution  of the Galactic center S-star orbits}

\author{Fabio Antonini}
\email{antonini@cita.utoronto.ca}
\affil{Canadian Institute for Theoretical Astrophysics, University of Toronto, 60 St. George Street, 
Toronto, Ontario M5S 3H8, Canada }
\author{David Merritt}
\email{merritt@astro.rit.edu}
\affil{Department of Physics and Center for Computational Relativity and Gravitation, Rochester Institute of Technology, 85 Lomb Memorial
Drive, Rochester, NY 14623, USA}

\begin{abstract}
We consider the orbital evolution of the S-stars, the young main-sequence 
stars near the supermassive black hole~(\sbh) at the Galactic center~(GC), and put constraints on  competing models for their origin.
Our analysis includes for the first time the joint effects of Newtonian and relativistic
perturbations to the motion, including the dragging of inertial frames by a spinning
\sbh\ as well as torques due to finite-$N$ asymmetries in the field-star distribution 
(resonant relaxation, RR). 
The evolution of the S-star orbits is strongly influenced by 
the Schwarzschild barrier~(SB),  the locus in the ($E,L$) plane where RR is ineffective
at driving orbits to higher eccentricities.
Formation  models that invoke tidal disruption of binary stars by the \sbh\
tend to place stars below (i.e., at higher eccentricities than) the SB;
some stars remain below the barrier, but most stars are
able to penetrate it, after which they are subject to RR and 
achieve a nearly thermal distribution of eccentricities.  
This process requires roughly $50~$Myr in nuclear models with relaxed stellar cusps,
or $\gtrsim10~$Myr, regardless of the initial distribution of eccentricities, 
in nuclear models that include a dense cluster of $10~M_{\odot}$  black holes.
We find a probability of $\lesssim 1~\%$ for  any  S-star to be tidally disrupted by the 
\sbh\ over its lifetime.
 \end{abstract}

\subjectheadings{
black hole physics-Galaxy:center-Galaxy:kinematics and dynamics-stellar
dynamics} 
%%%%%%%%%%%%%%%%%%%%%%%%%%%%%%%%%%%%%%%%%%%%%%%%%
\section{Introduction}
Observations of the Galactic center~(GC) reveal a cluster of about 20 stars, mainly 
main-sequence B stars, that extends outward to about a tenth of a parsec from the central 
supermassive black hole~\citep[\sbh;][]{Ghez2008}.
These stars, usually referred to as ``S-stars,'' follow orbits that are  randomly oriented and have a nearly ``thermal'' distribution of eccentricities, $N(e) de \sim e de$
 \citep{G2009}. 
The existence of such young
stars so close to the GC \sbh\ challenges our understanding of star formation since 
the strong tidal field of the \sbh\ should inhibit the collapse and fragmentation of 
molecular clouds~\citep{Morriss1993}. For this reason,
it is usually assumed  that the S-stars formed elsewhere and migrated 
to their current locations. However, the migration mechanisms proposed in the literature 
result in orbital distributions that differ substantially from what is
observed. Post-migration dynamical evolution due to
gravitational interactions with other stars or stellar black holes~(\bh s) has
been invoked to bring the predicted orbital distributions more in line with observations~\citep[e.g.,][]{Merritt+09,Perets2009,Madigan2011,ZLY12}.

The S-stars approach closely enough to Sgr A* that relativistic corrections to
their equations of motion can be important.  In this paper, we apply recent
insights about how relativity interacts with Newtonian (star-star) perturbations
near Schwarzschild and Kerr \sbhs. Using an approximate
Hamiltonian formulation that includes a post-Newtonian description of the
effects of relativity, we explore the evolution of the S-star orbits starting
from initial conditions that correspond to the different formation models
proposed in the literature.
Evolving the initial conditions for a time of the 
order the lifetime of the S-stars, and comparing with the observed distribution of orbital elements,  allows us to place constraints on both the parameters of the nuclear cusp and 
the S-star origin models. 

\section{Gravitatonal encounters near the \sbh}
%%%%%%%%%%%%%%%%%%%%%%%%%%%%%%%%%%%%%%%%%%%%%%%%%%%
\begin{figure*}
\centering
\includegraphics[width=.37\textwidth,angle=270]{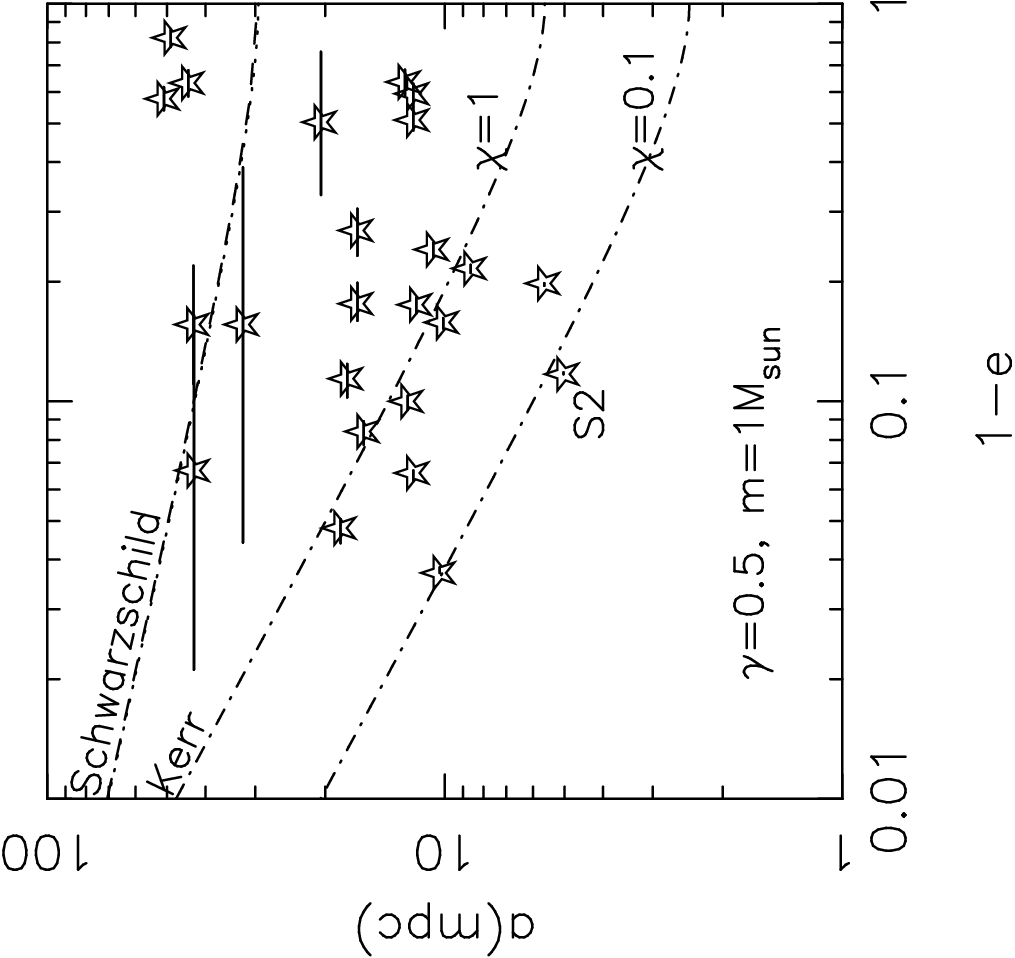}
\includegraphics[width=.37\textwidth,angle=270]{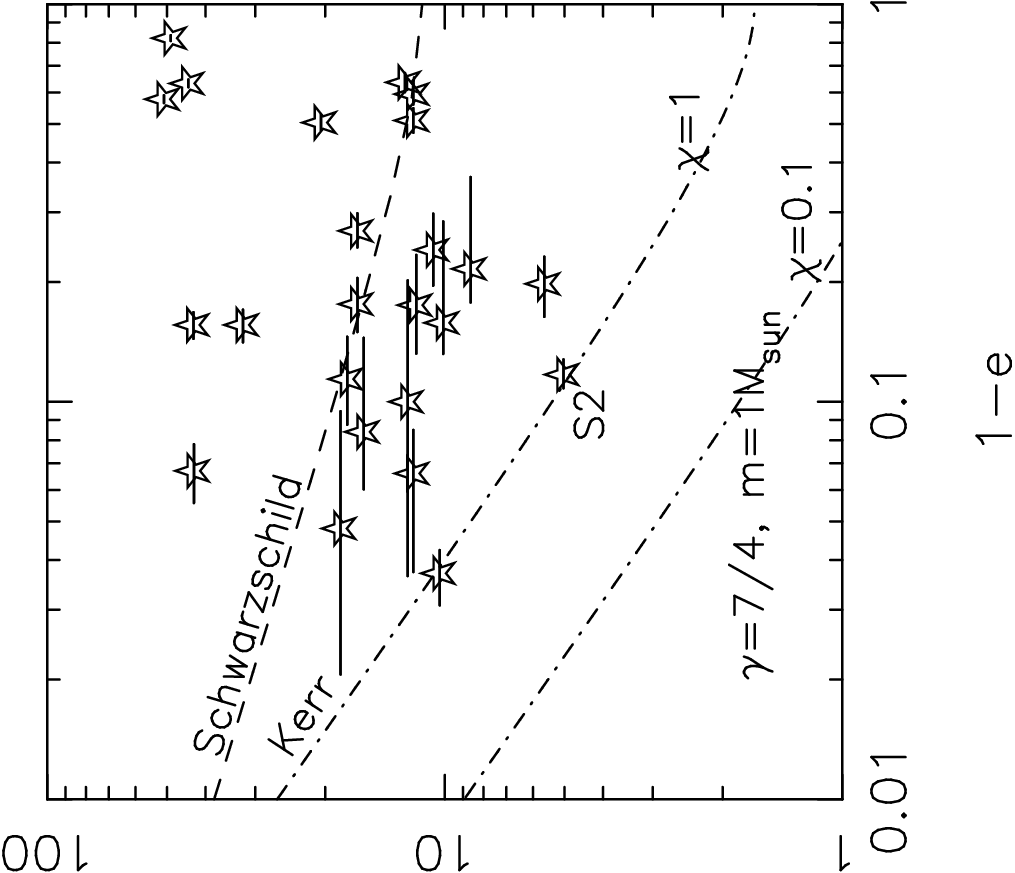}
\includegraphics[width=.37\textwidth,angle=270]{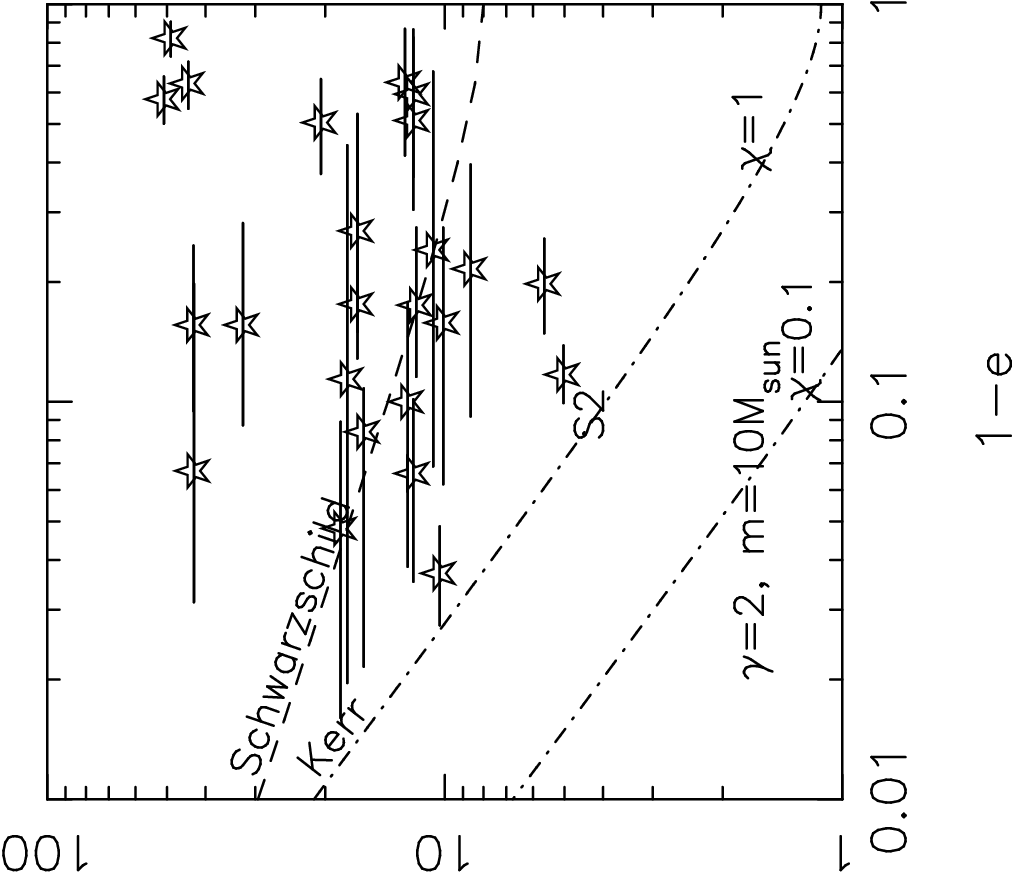}
\caption{Location of the Galactic center S-stars on the ($a,e$) plane,
compared with the Schwarzschild barrier (dashed line, equation~\ref{SB}),
and the curve along which frame-dragging torques compete with $\sqrt{N}$ torques
from the stars (dash-dotted lines, equation~\ref{Equation:aK}, with two different
values of the \sbh\ spin $\chi$).
The three panels are for three models of the nucleus, as described in the text.
The S-star data are from \citet{G2009}. 
Horizontal tick marks give the expected amplitude of eccentricity changes 
as an orbit precesses in the fixed torquing field due to the field stars~\citep[equation~41 from][]{MAMW2011}.
}\label{Fig1}
\end{figure*}

Timescales of interest are of order 100 Myr, the main-sequence lifetime of a B star, or less.
Such times are short compared with two-body (non-resonant, NR) relaxation times near the center of the Milky Way 
\citep[e.g.,][]{Merritt2010,AM2012}, hence we ignore NR relaxation in what follows and assume that orbital energies, i.e. semi-major axes $a$, are unchanged once
a star has been deposited near Sgr A$^*$.

Resonant relaxation (RR) \citep{RT1996,HA2006} acts to change
orbital eccentricities in a time
\begin{equation}
T_\mathrm{RR} = \left(\frac{L_c}{|\Delta L_\mathrm{coh}|}\right)^2 t_\mathrm{coh},
\end{equation}
the ``incoherent RR time,'' where $L_c$ is the angular momentum of a circular orbit having the same semi-major axis as the test star and $t_\mathrm{coh}$ is the ``coherence time,'' 
defined as the time for a typical field-star to change its orbital orientation; the latter is
the shortest of the mass precession time (due to the distributed mass), the relativistic
precession time (due to the 1PN corrections to the Newtonian equations of motion), 
and the time for
RR itself to reorient orbital planes.
For instance, in the case that field-star precession is dominated by relativity,
%\begin{equations}
\begin{eqnarray}\label{Equation:TRR1PN}
T_\mathrm{RR} &\approx& \frac{3}{\pi^2} \frac{r_g}{a} \left(\frac{\mh}{m}\right)^2\frac{P}{N(<a)} \\
&\approx& 1.4 \times 10^5 \left(\frac{a}{10\,\mathrm{mpc}}\right)^{1/2} 
\left(\frac{m}{1 M_\odot}\right)^{-2}
\left(\frac{N}{10^3}\right)^{-1} \mathrm{yr} \nonumber 
\end{eqnarray}
%\end{equations}
where $r_g\equiv G\mh/c^2$, $P$ is the orbital period of the test star, 
$N(<a)$ is the number of
field-stars with semi-major axes less than $a$, $m$ is the mass of the field stars, 
and mpc is milliparsecs;
$\mh=4\times 10^6 M_\odot$ has been assumed.

RR ceases to be effective at changing the eccentricities of  stars
whose orbits lie below (at higher eccentricities than) 
the ``Schwarzschild barrier'' (SB), the locus in the ($a,e$) plane
where relativistic precession of the test star acts in a time shorter than the time
for the field-star torques to change $L$.
The SB is defined approximately by \citep{MAMW2011}
\begin{eqnarray}\label{SB}
\left(1-e^2\right)_{\rm SB}^{1/2} \approx \frac{r_g}{a} \frac{\mh}{m}\frac{1}{\sqrt{N(<a)}} .
\end{eqnarray}
Orbits above (at lower $e$ than) the SB evolve in response to RR by undergoing a random
walk in $e$.
If such an orbit ``strikes'' the SB, it is ``reflected'' in a time
of order the coherence time and random-walks again to lower $e$, in a time $\sim T_\mathrm{RR}$, before eventually striking the SB again etc.
Penetration of the SB from above can occur but only on a timescale that is longer than
both the RR and NRR timescales \citep{MAMW2011}.

If a star should find itself {\it below} the SB, torques from the field stars are still able to change the orientation of its orbital plane (``2d RR'') even though changes in eccentricity are suppressed.
The timescale for changes in orientation is
\begin{eqnarray}\label{Equation:T2dRR}
T_\mathrm{2dRR} &\approx& \frac{P}{2\pi} \frac{\mh}{m_\star}\frac{1}{\sqrt{N(<a)}}\\
&\approx& 9.4\times 10^5 \left(\frac{a}{10 \mathrm{mpc}}\right)^{3/2} 
\left(\frac{m}{1 M_\odot}\right)^{-1}
\left(\frac{N}{10^3}\right)^{-1/2} \mathrm{yr}
\nonumber
\end{eqnarray}
where again $\mh=4\times 10^6 M_\odot$ has been assumed.
However, 2dRR itself ceases to be effective for orbits that come sufficiently
close to the \sbh, where dragging of inertial frames by a spinning \sbh\
  induces Lense-Thirring precession with a period that is shorter
than the time for 2dRR to randomize orbital planes. 
The condition for an orbit to be in this regime is \citep{MerrittVasiliev2012}
\begin{equation}\label{Equation:aK}
\left(1-e^2\right)^3 \left(\frac{a}{r_g}\right)^3 \lesssim \frac{16 \chi^2}{N(<a)}
\left(\frac{\mh}{m}\right)^2 
\end{equation}
with $\chi\equiv c S/(G\mh^2)$ the dimensionless spin of the \sbh, and
  $S$ the \sbh\ spin angular momentum.
We define $a_\mathrm{K}$,  the ``radius of rotational influence'' of the \sbh,
as the value of $a$ that satisfies equation (\ref{Equation:aK}) with $e=1$; 
$a_\mathrm{K}$ is roughly $1$ mpc  for the Milky Way assuming $\chi=1$ \citep{MAMW2010}.
%%%%%%%%%%%%%%%%%%%%%%%%%%%%%%%%%%%%%%%%%%%%%%%%%%
\begin{table*}\scriptsize
\begin{center}
\caption{Origin models for the S-stars \label{t1}} 
\begin{tabular}{lllllll}
 \hline
&&&~~~~~p$^a$ & & & TDs$^b$(\%)\\ 
   \cline{2-6}  \\
 Binary Disruption\\ Burst scenario$^c$ & 5Myr  & 20Myr  & 50Myr   & 100Myr &   200Myr \\
 \hline
$\gamma=0.5;~m=1M_{\odot}$      & $7.41\times 10^{-14}$ & $7.39\times 10^{-12}$ & $2.01\times 10^{-10}$ & $1.31\times 10^{-9}$ &  $9.36\times 10^{-9}$ &0 \\
 $\gamma=7/4;~m=1M_{\odot}$     & $5.11\times 10^{-5}$ & $0.176$  & $0.765$ & $7.93\times 10^{-2}$ &  
 $3.10\times 10^{-2}$  & 1.9\\
 $\gamma=2;~m=10M_{\odot}$      & $0.593$ & $0.296$  & $0.239$  &  $0.160$ & 0.201 & 0.36\\
\\ 
\hline
Migration from\\ gaseous  disk$^c$ &  5Myr   &20Myr  &50Myr & 100Myr &  200Myr	\\
 \hline
 $\gamma=0.5;~m=1M_{\odot}$      & $2.90\times 10^{-11}$ & $9.57\times 10^{-10}$ & $8.29\times 10^{-9}$ & $3.62\times 10^{-7}$ &  $1.12\times 10^{-5}$ & 0\\
 $\gamma=7/4;~m=1M_{\odot}$      & $3.12\times 10^{-9}$ & $6.14\times 10^{-7}$  & $4.06\times 10^{-5}$  & $1.70\times 10^{-4}$ &  $1.37\times 10^{-3}$  &0.12\\
 $\gamma=2;~m=10M_{\odot}$       &  $6.78\times 10^{-3}$  & $3.76\times 10^{-2}$  & $0.138$  &  $0.168$ &  $0.184$ &0.41\\
 \hline \\
 \hline
Binary Disruption \\ Continuous scenario$^d$ &  5Myr   &20Myr  &50Myr & 100Myr &  200Myr	\\
 \hline
 $\gamma=0.5;~m=1M_{\odot}$      & $7.21\times 10^{-13}$ & $6.41\times 10^{-13}$ & $5.32\times 10^{-13}$ & $1.33\times 10^{-13}$ &  $1.267\times 10^{-13}$ & 0\\
 $\gamma=7/4;~m=1M_{\odot}$      & $3.14\times 10^{-5}$ & $0.147$  & $0.645$  & $0.819$ &  $0.660$  & $0.16$\\
 $\gamma=2;~m=10M_{\odot}$       &  $8.07\times 10^{-2}$  & $0.108$  & $0.410$  &  $0.499$ &  $0.310$ &0\\
 \hline \\
\multicolumn{6}{l}{$^{a}${\footnotesize Probability value of the 2 samples
Kolmogorv-Smirnov test}}  \\
\multicolumn{6}{l}{$^{b}${\footnotesize Percentage of stellar tidal disruptions after $200~$Myr }} \\
\multicolumn{6}{l}{$^{c}${\footnotesize Orbits initialized at t=0}}\\
\multicolumn{6}{l}{$^{d}${\footnotesize Orbits initialized at random times between [0,200~Myr]}}
 \tabularnewline
\end{tabular}
	\end{center}
\end{table*}

While the joint evolution of an ensemble of stars near a spinning \sbh\ can only be
convincingly treated using an $N$-body code,
Monte-Carlo algorithms have been constructed that faithfully reproduce 
the eccentricity evolution of single (test) stars due to the dynamical mechanisms
described above, 
assuming that $N(a,e)$ for the field-star distribution is not evolving.
In this paper, we use an algorithm similar to that described by \citet{MAMW2011}.
The Hamiltonian that defines the test-star motion includes
terms representing the effects of the spherically-distributed mass (which results
in precession of the argument of periastron),  
1PN (Schwarzschild), 1.5PN (Lense-Thirring) precession due to relativity
and dipole and quadruple order terms representing the torquing due to the finite-$N$ asymmetry
in the field-star distribution (which induces changes in all the orbital elements).
The direction of the torquing field is changed smoothly with time and is randomized
in a time of $t_\mathrm{coh}$, as described in \S VB of \citet{MAMW2011}.

%%%%%%%%%%%%%%%%%%%%%%%%%%%%%%%%%%%%%%%%%%%%%%%%%%%%%%
\begin{figure*}
\centering
 \includegraphics[width=.3\textwidth,angle=270]{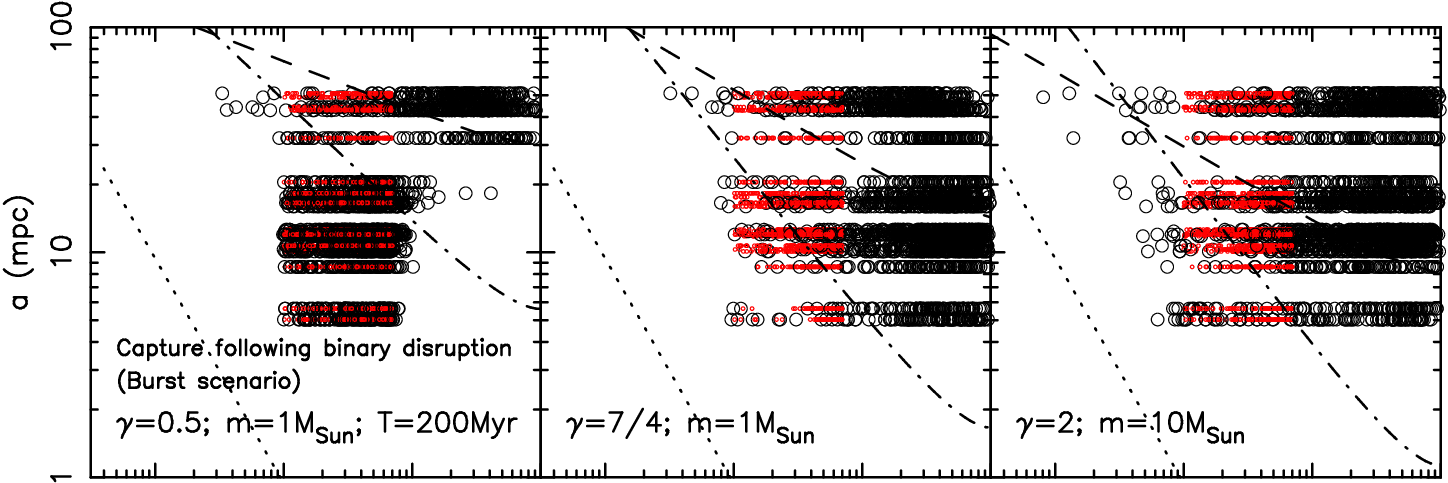}
  \includegraphics[width=.358\textwidth,angle=270]{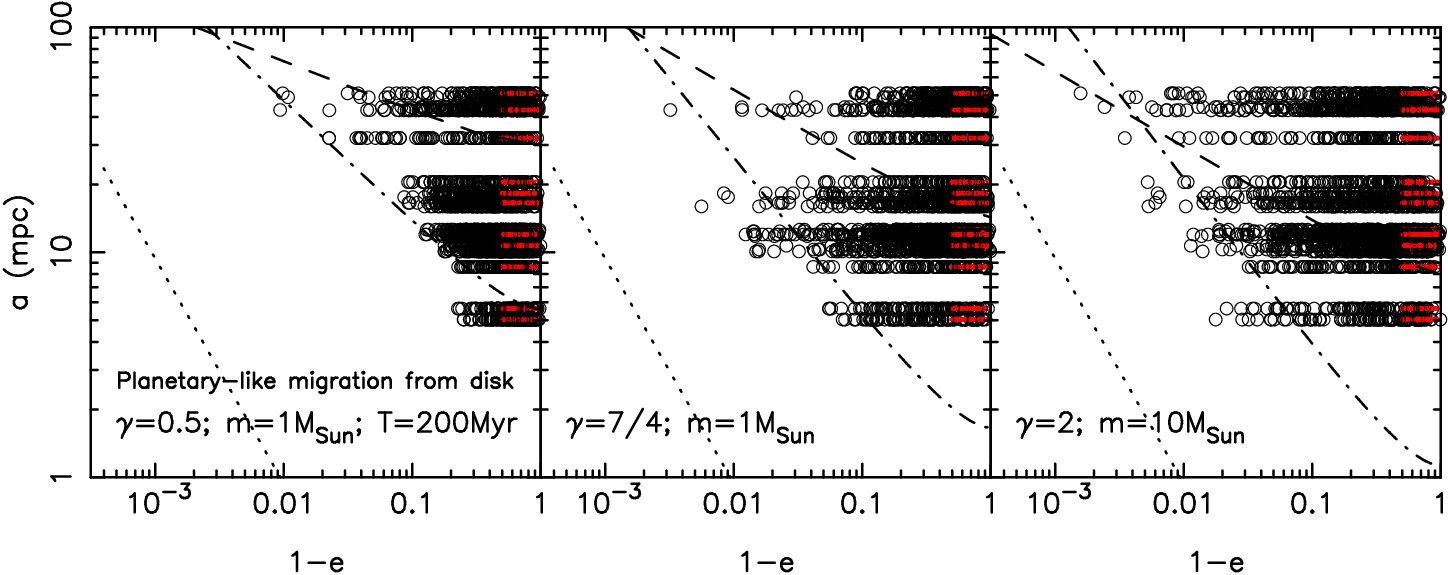}
\caption{Initial~(red points) and final~(black open circles, after 200 Myr) locations of  
of the test particles  in the Monte-Carlo integrations. 
Dashed and dot-dashed lines are defined in Figure~\ref{Fig1}; 
dotted lines give the  tidal disruption radius of a $10~M_\odot$ star.
Stars that initially have large eccentricities and lie near, but to the left of,  
the SB can penetrate the barrier and move to the right, where they remain. 
These stars end up with  a nearly uniform distribution of angular momenta.
Stars that  are initially right of the SB tend to remain there, 
though some barrier penetration (from right to left) is observed  near $a_\mathrm{crit}$, 
 the limiting value of $a$ for which the SB exists.
 }\label{Fig2} 
\end{figure*}

Above the SB, where test-star precession times are comparable with typical
field-star precession times, the assumptions underlying the derivation of
resonant relaxation are satisfied and the algorithm correctly reproduces the
eccentricity evolution predicted by RR, as well as the ``bounce'' observed in 
 $N$-body integrations when an orbit strikes the SB.
For a test star that finds itself {\it below }the SB, 
the Schwarzschild precession time is short compared
with field-star precession times.
In this regime, a test star precesses with period close to 
\begin{equation}
t_\mathrm{GR} = \frac{P}{3}\frac{a}{r_g}(1-e^2),
\end{equation}
the 1PN apsidal precession time.
During one precessional period, the field-star torques are nearly constant;
as a result, the test-star's angular momentum oscillates with period $t_\mathrm{GR}$ 
and with approximate amplitude (in the small-$\ell$ limit)
\begin{subequations}\label{Equation:DeltaL}
\begin{eqnarray}\label{Equation:DeltaLa}
\Delta \ell &\approx& 2 \ell_\mathrm{av}^2 A_\mathrm{D}\sin i ,\\
A_D &=& \frac{1}{3\sqrt{N(<a)}} \frac{M_\star(<a)}{\mh} \frac{a}{r_g} .
\label{Equation:DeltaLb}
\end{eqnarray}
\end{subequations}
\citep{MAMW2011}.
Here, $\ell^2=1-e^2$, $M_\star(<a)$ is the mass in stars at radii $r\le a$,
and $\sin i$ specifies the inclination of the major axis of the torquing potential with
respect to the orbit.
By themselves, these periodic variations in $e$ do not imply any 
directed evolution in angular momentum, but random switching 
of the direction of the torquing potential does result in a random walk in a test star's angular momentum, 
allowing a star that is initially below the SB to approach it.

Dragging of inertial frames results in orbit-averaged rates of change of the argument 
of periastron, $\omega$, and the angle of nodes, $\Omega$, of the test star according to:
\begin{subequations}\label{Equation:Omegaomega}
\begin{eqnarray}
\left( \frac{d\Omega}{dt} \right)_{\rm FD} &=& \frac{2G^2\mh^2}{c^3a^3(1-e^2)} \chi\,  ,\\
\left( \frac{d\omega}{dt} \right)_{\rm FD} &=& -\frac{6G^2\mh^2}{c^3a^3(1-e^2)} \cos i\,  \chi .
\end{eqnarray}
\end{subequations}
In equations (\ref{Equation:Omegaomega}), the ``reference plane'' for
$\Omega$ and for $i$ (the orbital inclination) is the \sbh\ equatorial plane.

Since little is known about the distribution of stars and stellar remnants 
near the Galactic center,
we  explored a range of different models for the field-star distribution.
Assuming a power-law density profile, the number of stars at radii less than $r$ 
is
\begin{equation}
N(<r)=  N_{0.2}\left(\frac{r}{0.2~{\rm pc}}\right)^{3-\gamma}~,
 \end{equation}
 where $N_{0.2}\equiv N(<0.2~{\rm pc})$.
 The  parameters $\{N_{0.2},\gamma, m\}$  then uniquely 
define the background  distribution in which the test particle orbits are evolved.
In two models, we set $m=1~M_\odot$,  
and we take   either
$\gamma=0.5$ and $N_{0.2}=8\times 10^4$~\citep[equation (5.247) in][]{M2012},  
similar to what is inferred from observations~\citep[e.g.,][]{Do2009},
or $\gamma=7/4$ and $N_{0.2}=1.6\times 10^5$~\citep[equation (5.246) in][]{M2012}, 
 the expected values for a dynamically-relaxed population of stars~\citep{BW1977}.
In another model, we adopt $N_{0.2}=4.8\times10^3,~\gamma=2$ and $m=10M_\odot$.
This latter choice of parameters approximately reproduces the
density of stellar BHs  predicted by collisionally relaxed models of a cusp of stars and 
stellar remnants around SgrA*~\citep{HA2006}.

Figure~\ref{Fig1} plots the S stars on the ($a,e$) plane, as well as the location of the SB;
 the latter depends on the parameters defining the nuclear cusp through equation (\ref{SB}).
 Dot-dashed lines in the figure delineate the region where frame-dragging torques from
a spinning \sbh\ would dominate stellar torques, equation~(\ref{Equation:T2dRR}).
Particularly for large $\chi$, many of the S stars lie close to this transition region, 
suggesting that frame dragging could be an important influence on their orbital evolution;
for instance, by inhibiting 2dRR. 
This figure does not give information about timescales, but we note that characteristic
times like $T_\mathrm{RR}$ are functions of the nuclear parameters, which can be important
given the limited lifetimes of the S stars.

In the  two models with a steep cusp, 
the SB  delineates the boundary of the  S-star orbits, 
with only a few stars lying below the minimum $a=a_\mathrm{crit}\approx (\mh/m\sqrt{N})r_g$ for which the barrier exists. In these models, the existence of the SB is expected to strongly influence the
eccentricity evolution of at least some of the S stars; for instance, by limiting the maximum
eccentricity attainable by a star that starts above the barrier.
Furthermore, for some of the nuclear models, 
Figure~\ref{Fig1} shows that some of the S-stars can lie both above 
$a_\mathrm{crit}$ and below the SB.
Such a location would be highly unlikely, in a time as short as $\sim100~$Myr, for stars
that started above the SB, but is reasonable 
if the stars were placed initially on such orbits via one of the
mechanisms described below.
%%%%%%%%%%%%%%%%%%%%%%%%%%%%%%%%%%%%%%%%%%%%%%%%
\section{Formation models}
We considered two models for the origin of the S stars. 
\begin{itemize}
\item[(1)]
Formation of the S-stars in binaries far from the center~($r>0.1~$pc).
In this model, the binaries are scattered onto low-angular-momentum orbits
that bring them close enough to the \sbh\ that an exchange interaction 
can occur, leaving one star on a tightly-bound orbit around SgrA*~\citep{Hills1988,YT2003,ant+10}.
The radius at which the \sbh\ tidally disrupt a binary is
 typically a few tens of AU for main sequence binaries, and
the orbital eccentricity is expected to be large.
The initial orbital inclinations of the  S-stars 
will be either randomly distributed  if the binaries originated 
in an isotropic stellar cusp~\citep{Perets2007,PG10} or  highly 
correlated if they formed in a stellar disk~\citep{Madigan2009}.  
\item[(2)] Formation of the S-stars in a disk at roughly their current radius,
either one of the known stellar disks, or a pre-existing one. 
Formation in the disk would be followed by migration to their current locations.
This model predicts initially small eccentricities and inclinations~\citep{Levin2007}.
\end{itemize}
%%%%%%%%%%%%%%%%%%%%%%%%%%%%%%%%%%%%%%%%%%%%%% 
\begin{figure*}
\centering
$\begin{array}{lll}
\includegraphics[width=.27\textwidth,angle=0]{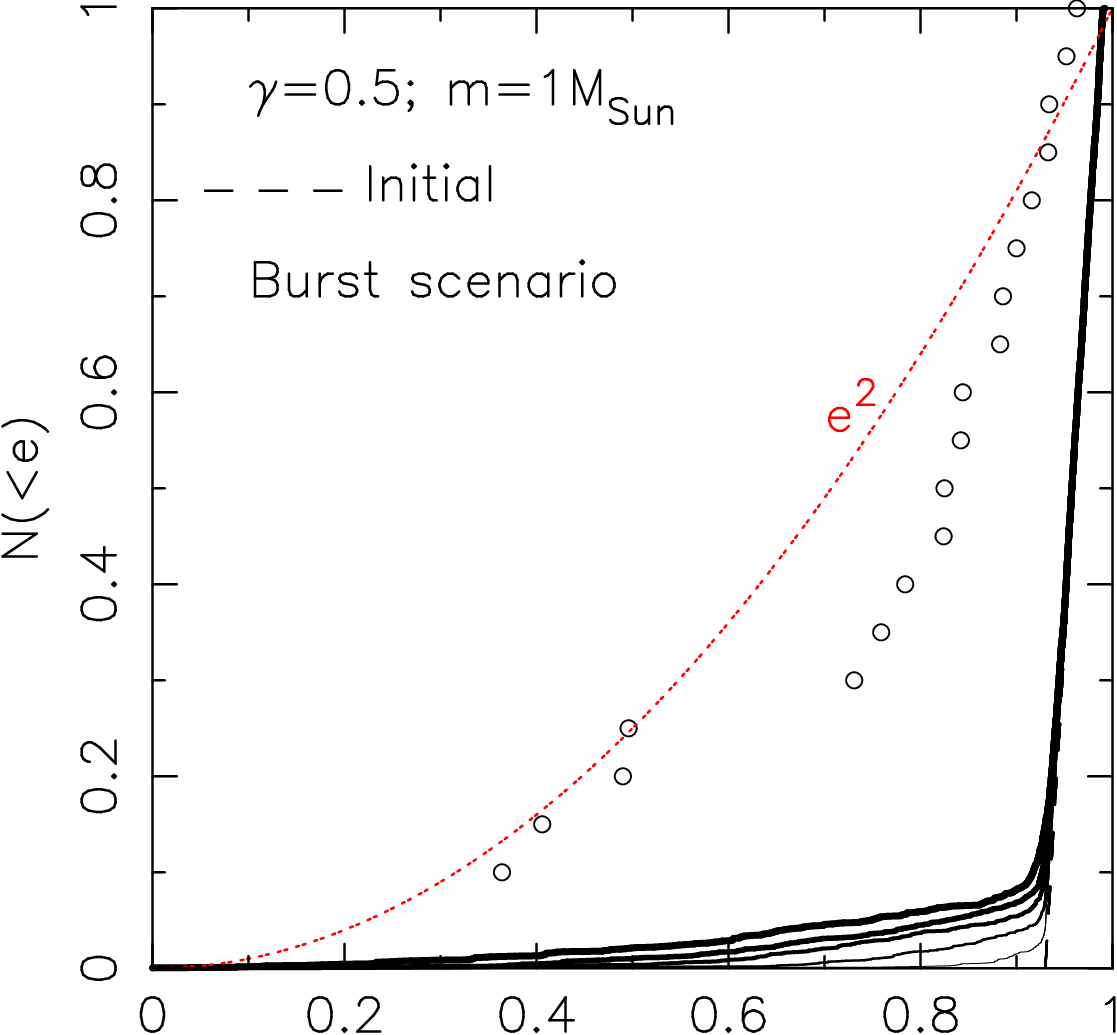} & 
\includegraphics[width=.25\textwidth,angle=0]{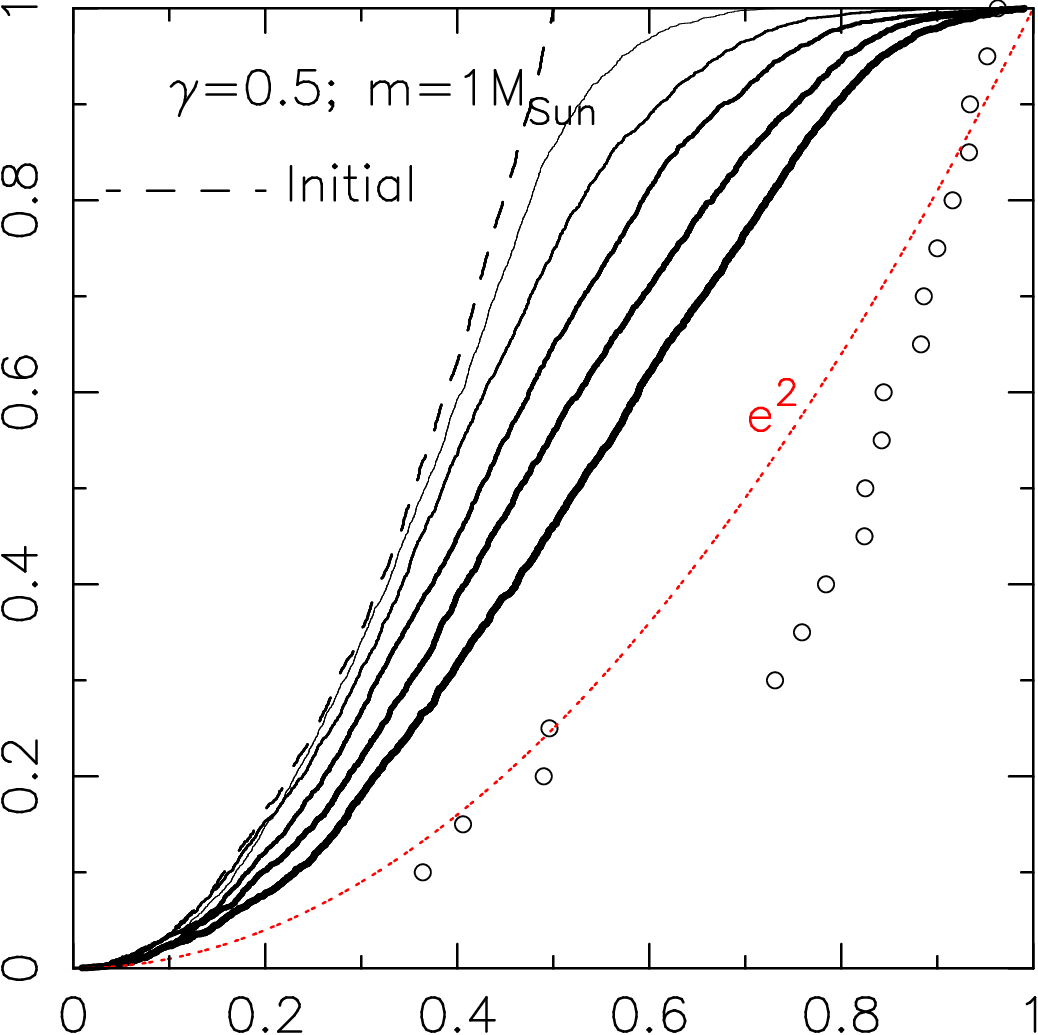} &
\includegraphics[width=.25\textwidth,angle=0]{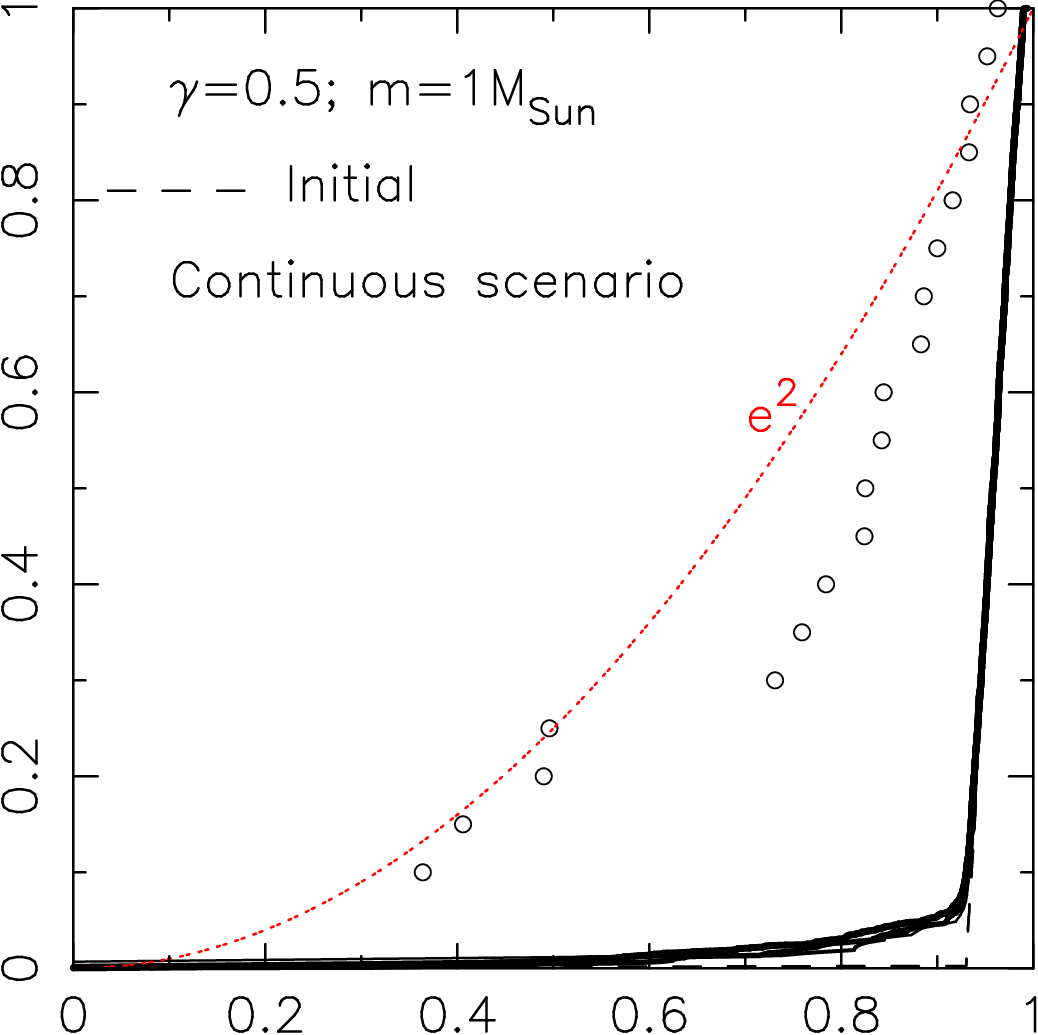} \\
\includegraphics[width=.27\textwidth,angle=0]{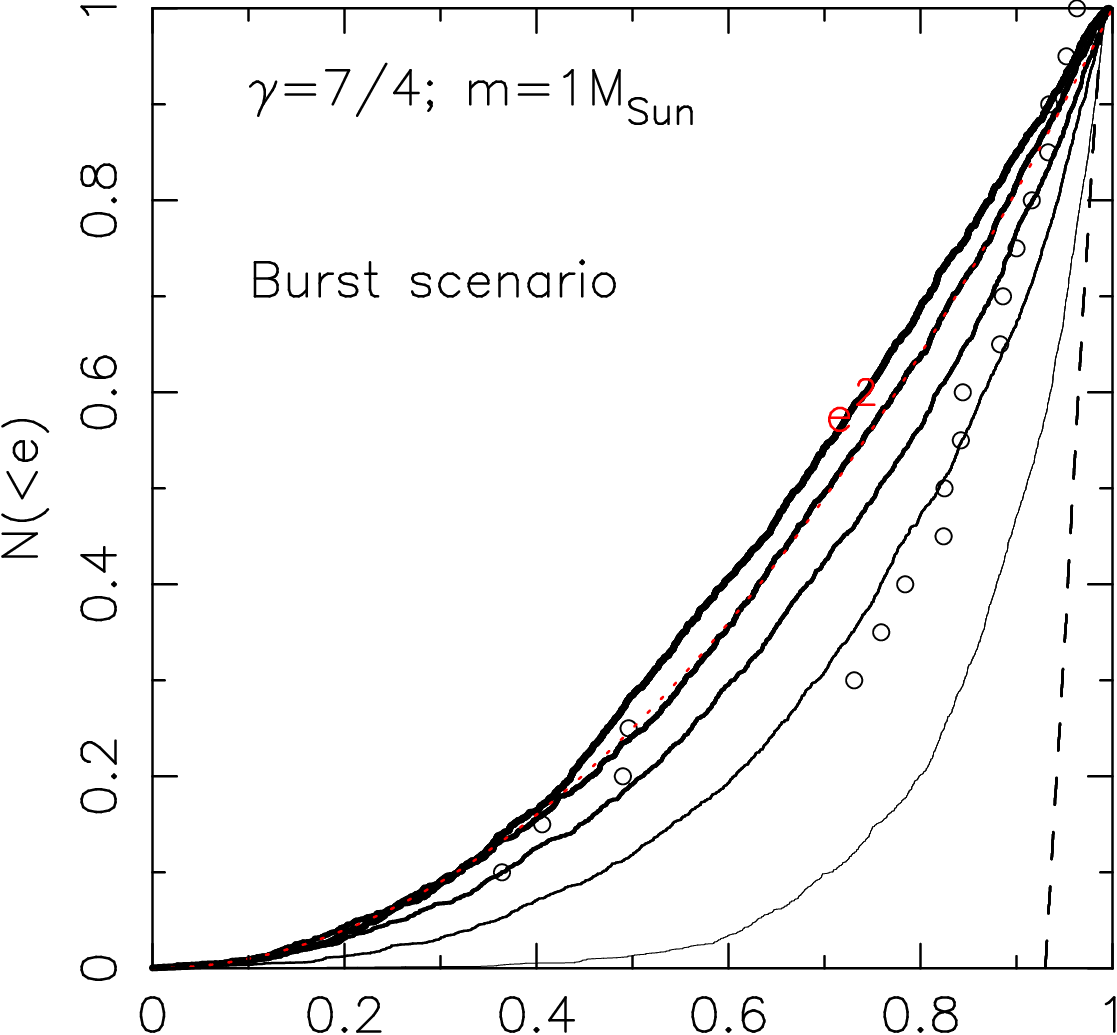} &
\includegraphics[width=.25\textwidth,angle=0]{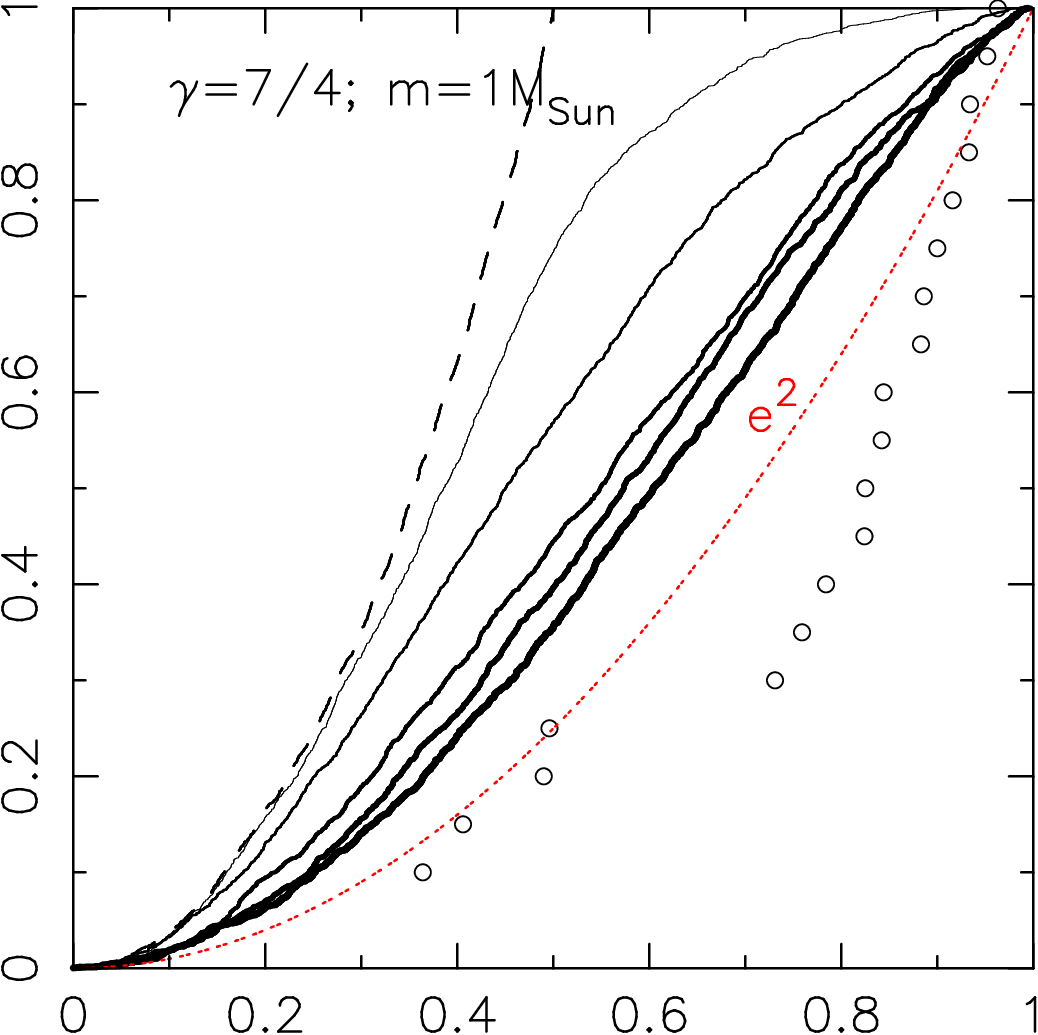} &
\includegraphics[width=.25\textwidth,angle=0]{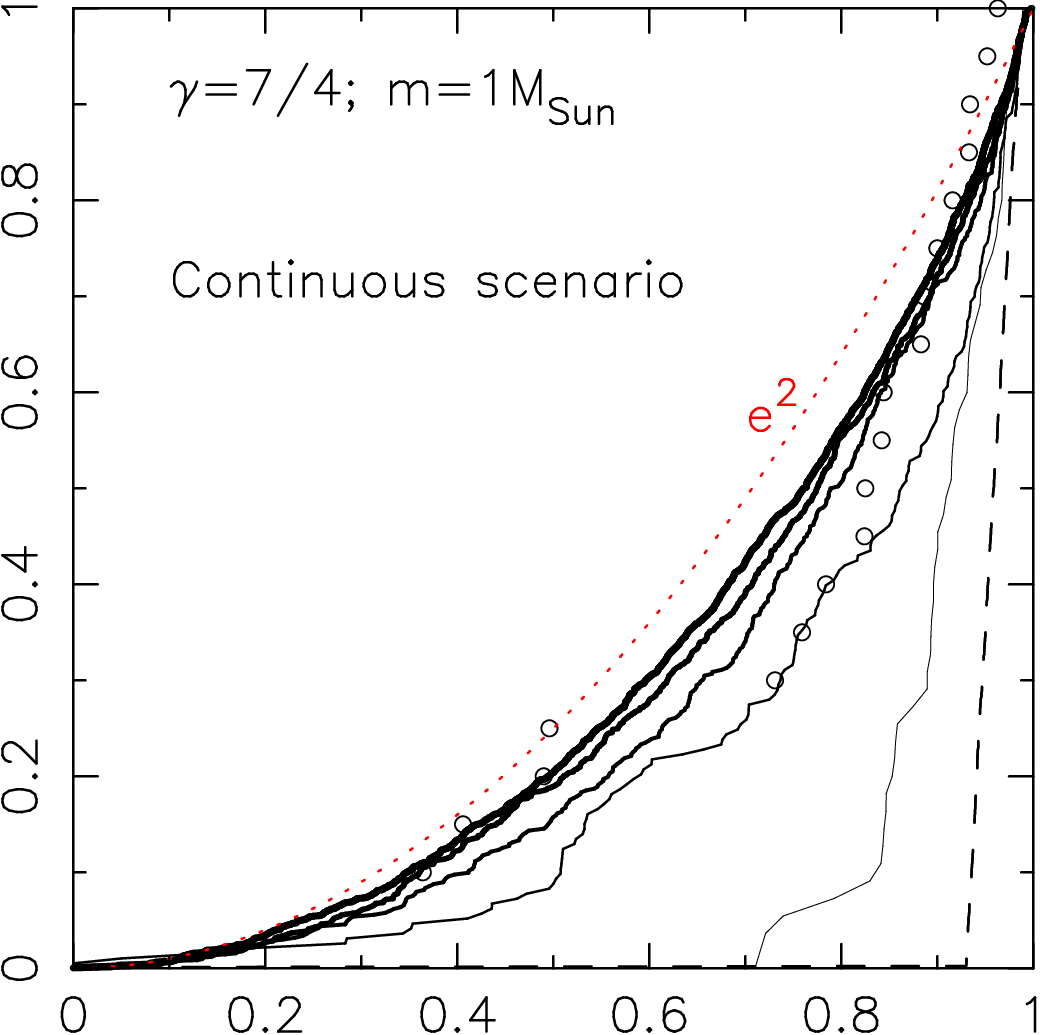} \\
\includegraphics[width=.27\textwidth,angle=0]{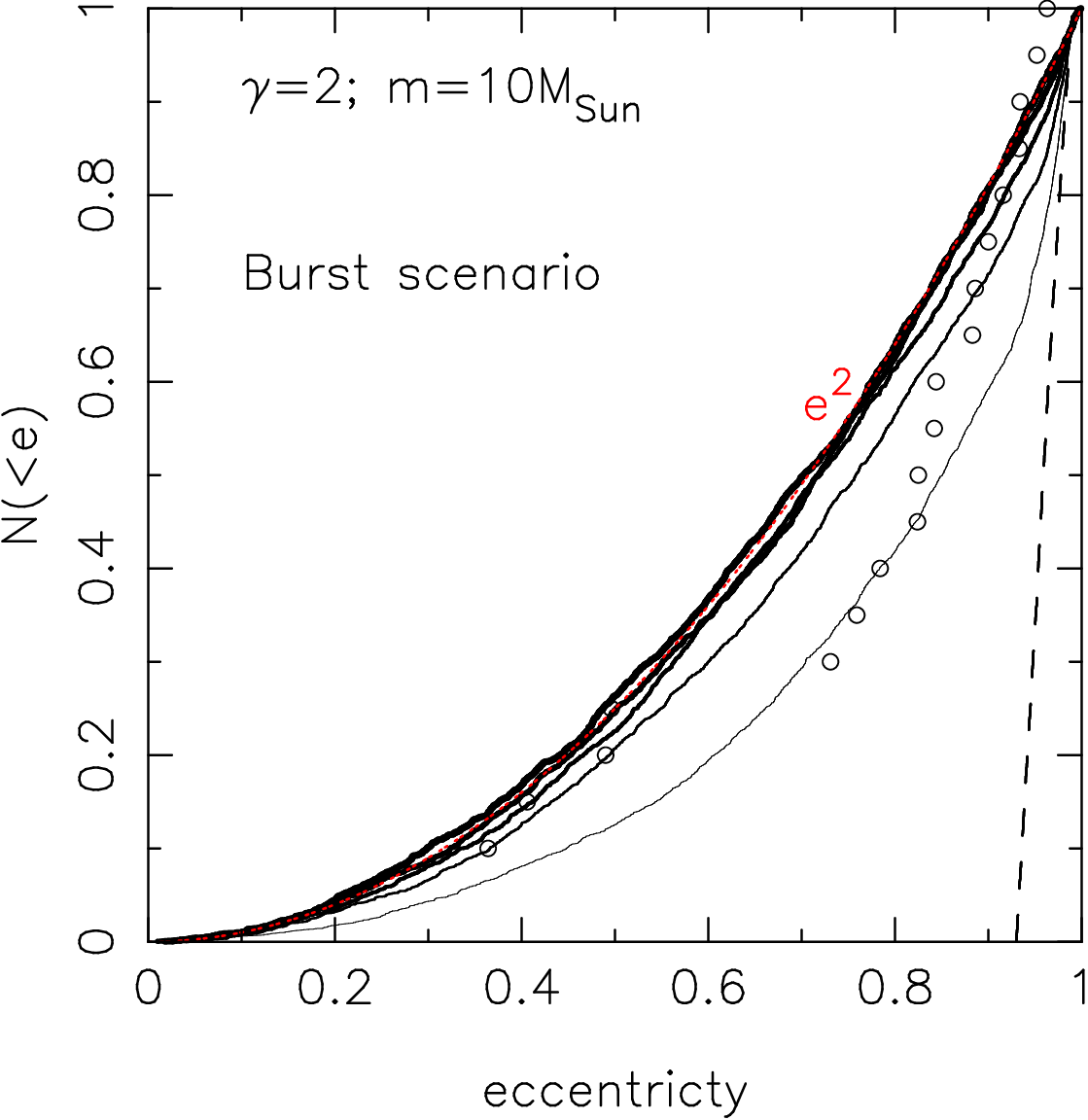} &
\includegraphics[width=.25\textwidth,angle=0]{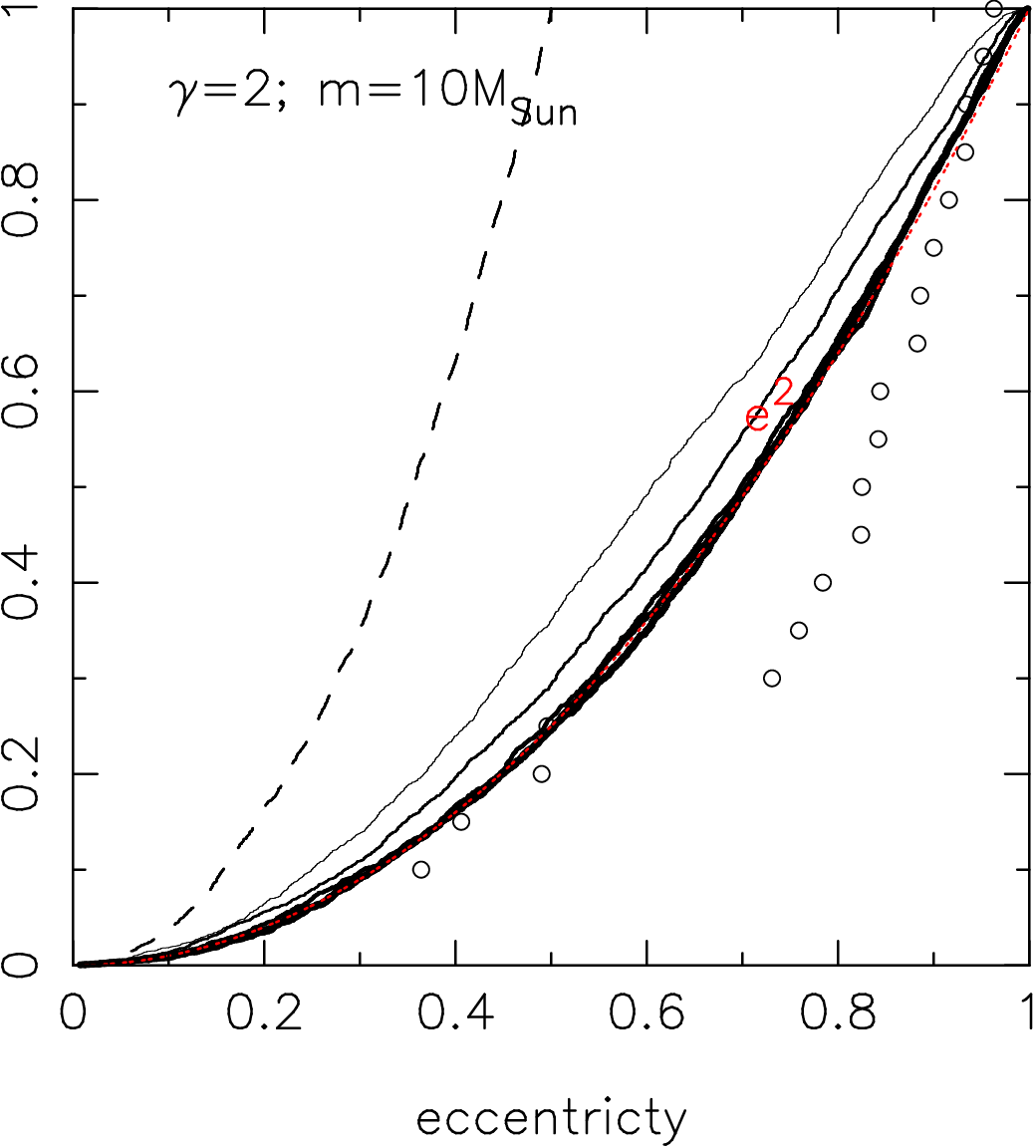} &
\includegraphics[width=.25\textwidth,angle=0]{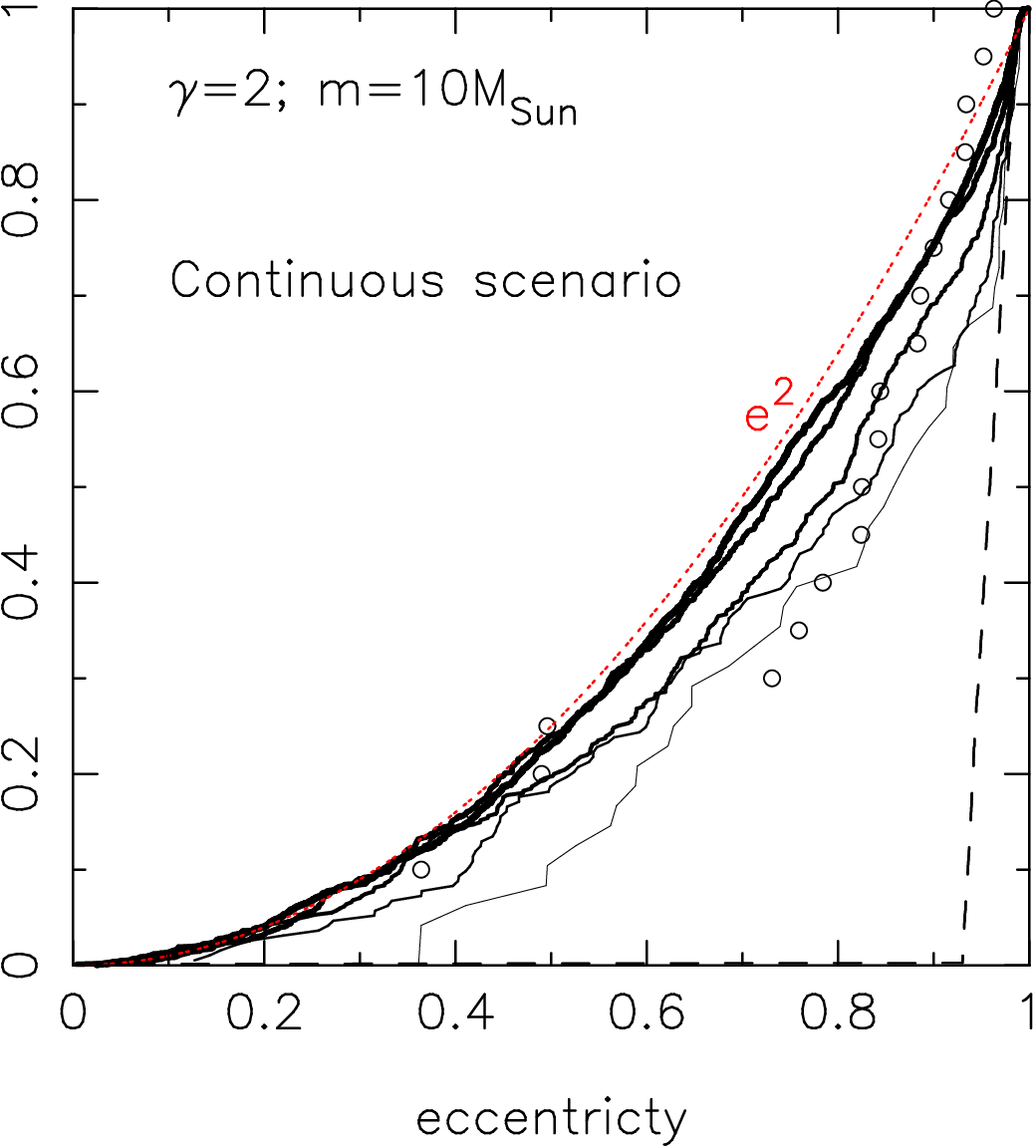} 
\end{array}$
\caption{Cumulative distribution of eccentricities for different S-star 
formation models after~$5,~20,~50,~100$ and $200$Myr (line thickness increases with time). 
Open circles give the observed distribution; dot-red line shows a ``thermal''  eccentricity distribution.
Left and right panels correspond to formation through capture following binary disruption, middle panels 
to formation in a stellar disk followed by migration through interaction with a gaseous disk.\\
}\label{Fig3}
\end{figure*}

\section{Orbital Evolution}

We start  by assuming that $N(a)$ is known: it is given by the observed values of $a$.
For each  S-star~(i.e. for each value of $a$), 100 Monte-Carlo experiments were carried
out using the Hamiltonian model described above, in each of the three nuclear models,
for an integration time of $200~$Myr. 
The initial orbital eccentricities were assigned from a thermal distribution, $N(<e)\propto e^2$,
over some specified range in $e$.
In the case of  migration from  a gaseous disk we required $e \leq 0.5$ initially. 
When considering the binary disruption model, we 
only considered orbits with initial eccentricities in the range $0.93\leq e \leq0.99$.
In this case, we assume either that the S-stars are brought to their current location at the same 
time~(\emph{burst scenario}) or that they arrive at random times between $0$ and $200~$Myr~(\emph{continuous scenario}).
The former  choice corresponds to a burst of S-star formation, for instance in a stellar disk~\citep{Madigan2009}, 
while the latter choice assumes that the S-stars form continuously in the isotropic stellar cusp~\citep{Perets2007}.  In all cases we set $\chi=1$. 

Stars were assumed to be tidally disrupted when they approached  the \sbh\ within 
 a distance $r_t=2R \left(M_\bullet/m \right)^{1/3}$~\citep{ALM2011}, with $m=10~M_{\odot}$ and $R=8~R_{\odot}$.

Figure~\ref{Fig2} compares the initial and  final (after~$200$~Myr) $a-e$ 
distributions. 
 In the two, steep-cusp models with high initial $e$, most of the stars start off
 to the left of the SB, where evolution in angular momentum is strongly suppressed by
the Schwarzschild precession.
Nevertheless, it can be seen that, after some time,  
an initially eccentric population separates into two subpopulations: 
stars that are so far leftward of the SB initially that their eccentricities hardly evolve;
and stars that either begin rightward of the SB, and remain there, or that are close enough 
initially to the barrier to cross it. 
The latter stars are subject to RR after crossing the barrier and end up with a nearly thermal eccentricity distribution. 
A clear ``gap'' between the two populations is evident in several of the frames of Figure~\ref{Fig2}; 
the gap extends from the SB on the right, to a somewhat higher eccentricity on the left. 
The SB acts like a membrane that is permeable in one direction only, from
left to right: having crossed the barrier from left to right, a star moves quickly (in a time of
$\sim T_\mathrm{RR}$) to a region of lower $e$ where 
it remains~\footnote{This behavior has been confirmed via direct $N$-body integrations.}.
The observed  evolution below the SB 
is due to  reorientation  of the torquing potential which results in a random walk in angular momentum,
 preferentially  toward lower $e$~\citep[see  \S{VB} of][]{MAMW2011};
this mechanism is qualitatively similar to resonant relaxation but obeys a different 
set of relations \citep{AlexanderMerritt2012}. 

On the other hand, when eccentricities are initially low, as in the bottom panels
of Figure~\ref{Fig2}, they tend to remain low, i.e., above the SB.
As noted in \citet{MAMW2011}, the permeability of the SB tends to increase
 near $a_\mathrm{crit}$ and this can be seen in Figure~\ref{Fig2} as well.
%%%%%%%%%%%%%%%%%%%%%%%%%%%%%%%%%%%%%%%%%%%%%%%%%%%%%%%%

Time evolution of the cumulative distribution of  eccentricities is shown in Figure~\ref{Fig3}.
(In this figure and in the analysis that follows, we do not include   S-stars that  likely belong to the disk(s) of O/WR 
stars, i.e.~S66,~S67,~S83,~S87,~S96,~S97).
Eccentricity distributions  were found to approach a nearly ``thermal" form 
in a time of order $T_{\rm RR}$. 
Based on  Figure~\ref{Fig3}, we see that
the lifetime of the S-stars  may, or may not, be long enough  
for this to happen, depending on the nuclear model.
We  compared the results of the Monte-Carlo integrations with S-star data by performing 
 2-sample Kolmogorov-Simornov~(K-S) tests on the $e$ distributions~(Table~\ref{t1}).
The best match to observations is attained after $20~$Myr of evolution in \emph{stellar} cusp models in the continuous 
scenario with $\gamma=7/4$ and starting from initially high eccentricities~(K-S test $p$-values of $\approx 0.7$).
Integrations that include stellar \bh s  also generate 
 orbital distributions which are in agreement with observations after approximately
 $5$ and $20~$Myr of evolution  for initially high and low eccentricity distributions respectively.

We  tested the degree of randomness of the orbital planes using the Rayleigh 
statistic~$\mathcal{R}$ \citep{RA1919}, defined as the resultant of the unit vectors $l_i,~i=1...N_{\rm mc}$, where 
$l_i$ is perpendicular to the orbital plane of the $i_{th}$ star and $N_{\rm mc}$ is the total number of
Monte-Carlo data points~(i.e.,~$N_{\rm mc}=1900$). Since the test stars were initialized with the same inclination,
the orbits are initially strongly correlated and  $\mathcal{R} \approx N_{\rm mc}$;
over a time of order $T_\mathrm{2dRR}$, 2dRR tends to randomize the orbital planes
and  $\mathcal{R}$ approaches  $\sqrt{N_{\rm mc}}$, the value  expected  for a random distribution.
When the main contribution to  dynamical relaxation comes from 
 stellar \bh s,
 $\mathcal{R}$ reached values consistent 
with isotropy after  $\sim20~$ and $\sim50~$Myr for initially high and low eccentricity distributions  respectively. 
In the \emph{stellar}-cusp model with $\gamma=7/4$
and staring from initial high $e$, $\mathcal{R}$ reached values consistent 
with isotropy after $\sim100~$Myr, while in all the other stellar-cusp models  we measured  
a significant departure from  randomness  ($\mathcal{R}/N_{\rm mc}\gtrsim0.2$)  even after $200~$Myr of evolution.
We estimated $\mathcal{R}$  separately for orbits that  any time were below  the dot-dashed lines of Figure~\ref{Fig2}
and found that this population had a distribution of orbital planes which was  less consistent 
with being random. Evidently,   2dRR was somewhat inhibited by frame dragging for orbits with initially large eccentricities.

The fraction of stars that would have been tidally disrupted after $200~$Myr of evolution
 was never larger than $\sim 1\%$~(Table~\ref{t1}).  As a comparison, in purely Newtonian integrations,
 \citet{Perets2009} found that up to $\sim 30~$\% of stars were disrupted after $20~$Myr of evolution.

%%%%%%%%%%%%%%%%%%%%%%%%%%%%%%%%%%%%%%%%%%%%%%%%%%
 \section{Conclusions}
In this paper we  studied the combined effects of Newtonian and relativistic
perturbations on the angular momentum evolution of the Galactic center~(GC) S-stars.
For the first time we have shown that the $a-e$ distribution of the S-stars predicted by the binary disruption model,
in which the stars are delivered to the GC on high-eccentricity orbits, 
is  consistent with the observed orbits even when relativistic effects are considered. 
In these formation models, most of the orbits lie initially below the 
Schwarzschild barrier, the locus in the (a,e) plane where resonant relaxation is ineffective at changing eccentricities.
Contrary to this basic prediction, we found that orbits starting sufficiently close to the barrier are 
sometimes able to penetrate it, diffusing above and reaching a nearly thermal e-distribution;
a small fraction of stars  remain confined below the barrier at low angular momenta~($e \gtrsim 0.95$).
A good match to observations is achieved  after $\sim20$Myr of evolution if the distribution
 of field stars at the GC follows a nearly dynamically relaxed form.
 Models that include a mass-segregated population of stellar \bh s  also generate after $\sim10~$Myr of evolution 
 distributions that are marginally consistent with  observations. 
 
 Based on the origin models considered here, the S-star obits can only be reproduced by 
 postulating dynamically relaxed states  (i.e., steep density cusps) for the GC. 
 This result is interesting given that such models are currently disfavored by observations~\citep{BSE,Do2009} and by some theoretical 
 arguments~\citep{Merritt2010,ant+11,gm+12}.

 \bigskip
 DM was supported in part by the National Science Foundation under grant no. 08-21141 and by the National Aeronautics and Space Administration under grant no. NNX-07AH15G.

%\begin{figure*}
%\centering
%\includegraphics[width=.33\textwidth,angle=270]{./CT/CT0.ps}
%\includegraphics[width=.33\textwidth,angle=270]{./CT/CT1.ps}
%\includegraphics[width=.33\textwidth,angle=270]{./CT/CT2.ps}
%\includegraphics[width=.33\textwidth,angle=270]{./TS/TS0.ps}
%\includegraphics[width=.33\textwidth,angle=270]{./TS/TS02.ps}
%\includegraphics[width=.33\textwidth,angle=270]{./TS/TS02-2.ps}
%\caption{}\label{Fig1}
%\end{figure*}

\end{document}